\documentclass{amsart}

\usepackage{chapter-rna-branching}

\usepackage[T1]{fontenc}
\usepackage{lmodern}
\usepackage{microtype}

\usepackage{tikz,pgfplots}
\usepackage{tkz-euclide}
\usetkzobj{all}
\usetikzlibrary{arrows, calc, backgrounds}
\tikzset{
  lppoint/.style={fill=black, circle, minimum size = 2mm, inner sep = 0mm},
  llppoint/.style={fill=black, fill opacity = .2, circle, minimum size = 2mm, inner sep = 0mm},
  fcentry/.style={draw,rounded corners=5pt,rectangle,text opacity=1,align=center},
  fcarrow/.style={->, very thick},
  fcwrapper/.style={rounded corners=1cm, fill=black, fill opacity=.2},
  cone1/.style={fill = black, fill opacity = 0.6},
  cone2/.style={fill = black, fill opacity = 0.2},
  cone3/.style={fill opacity = 0.8},
  cone4/.style={fill = black, fill opacity = 0.4},
  polyfill/.style={fill = black, fill opacity = 0.1},
}

\newcommand{\vecname}[1]{\ensuremath{\mathbf{#1}}}

\newcommand{\codeletter}[1]{\texttt{#1}}

\begin{document}
\title[Branching polytopes for RNA sequences]{Geometric combinatorics and computational molecular biology: Branching polytopes for RNA sequences}

\author[Drellich]{Elizabeth Drellich}
\address{Department of Mathematics, University of North Texas, Denton, TX, USA 76203}
\email{elizabeth.drellich@unt.edu}

\author[Gainer-Dewar]{Andrew Gainer-Dewar}
\address{Department of Mathematics and Computer Science, Hobart and William Smith Colleges, Geneva, New York, USA 14456}
\curraddr{Center for Quantitative Medicine, UConn Health, Farmington, CT, USA 06051}
\email{andrew.gainer.dewar@gmail.com}

\author[Harrington]{Heather A.~Harrington}
\address{Mathematical Institute, University of Oxford, Andrew Wiles Building, Radcliffe Observatory Quarter, Woodstock Road, Oxford OX2 6GG United Kingdom}
\email{harrington@maths.ox.ac.uk}

\author[He]{Qijun He}
\address{Department of Mathematical Sciences, Clemson University, Clemson, SC, USA 29634}
\email{qhe@clemson.edu}

\author[Heitsch]{Christine Heitsch}
\address{School of Mathematics, Georgia Institute of Technology, Atlanta, GA, USA 30332}
\email{heitsch@math.gatech.edu}

\author[Poznanovi\'{c}]{Svetlana Poznanovi\'{c}}
\address{Department of Mathematical Sciences, Clemson University, Clemson, SC, USA 29634}
\email{spoznan@clemson.edu}

% TODO: Do we have a grant number or anything specific we should say here?
\thanks{
  The authors' collaboration was funded by the Mathematics Research Communities program of the AMS.
  S.~Poznanovi\'c was also supported in part by NSF grant DMS-1312817. HAH gratefully acknowledges the support of EPSRC Fellowship EP/K041096/1.
  C.~Heitsch was supported in part by a Career Award at the Scientific Interface from the Burroughs Wellcome Fund.
}

\maketitle

\begin{abstract}
  Questions in computational molecular biology generate various discrete optimization problems, such as DNA sequence alignment and RNA secondary structure prediction.  However, the optimal solutions are fundamentally dependent on the parameters used in the objective functions.  The goal of a parametric analysis is to elucidate such dependencies, especially as they pertain to the accuracy and robustness of the optimal solutions.  Techniques from geometric combinatorics, including polytopes and their normal fans, have been used previously to give parametric analyses of simple models for DNA sequence alignment and RNA branching configurations.  Here, we present a new
  computational framework, and proof-of-principle results, which give the first complete parametric analysis of the branching portion of the nearest neighbor thermodynamic model for secondary structure prediction  for real RNA sequences.
\end{abstract}

\section{Motivation}
\label{s:motivation}
Over the past four decades, improvements in biotechnology have greatly accelerated the amount of biological sequence data available.
Yet, a fundamental challenge in computational molecular biology remains to reliably infer functional information from the linear encoding of DNA, RNA, and protein molecules.

As articulated in the central dogma of molecular biology, genetic information is stored in DNA from which it is transcribed into messenger RNA, and then translated into proteins by ribosomal and transfer RNA.
However, as always in biology, a wealth of complexity lurks below the surface of this basic principle.
Historically, most interest was focused on DNA sequences (as the cellular genome) and protein structures (as the cellular machinery).
Since the early 2000's, though, attention has increasingly turned to RNA as many more critical functions have been revealed, including gene splicing, editing, and regulation.

Like DNA, RNA is a sequence of nucleic acids, abbreviated \codeletter{A}, \codeletter{C}, \codeletter{G}, and \codeletter{U} (instead of \codeletter{T}), which form the familiar Watson-Crick pairings.
Unlike the canonical double-stranded DNA helix, most RNA molecules are naturally single-stranded and the intra-sequence base pairings are an integral component of the three-dimensional structure.
This is in contrast to the more subtle amino acid interactions which govern the formation of protein structures.
However, knowing the structure of a protein or RNA molecule is critical to understanding and manipulating its cellular functions.

The structure of an RNA molecule is understood hierarchically, from the linear biochemical chain through the planar set of canonical base pairings to the 3D molecule, whose structure includes more complicated tertiary interactions like pseudoknots and base triples.
Given the prevalence of RNA sequence data and the difficulties, both experimental and computational, of determining 3D structures, the majority of attention has focused on the set of base pairs, known as the \emph{secondary structure}.
In particular, one seeks to answer: \textit{What are the native base pairs for a given RNA sequence?}
As explained below, it is possible to efficiently compute an optimal secondary structure under the current Nearest Neighbor Thermodynamic Model (NNTM).
However, when these minimum free energy (MFE) predictions are compared to structures derived from information-theoretic means, the current gold-standard, the average accuracy for longer ribosomal RNA sequences is only 40\%.
Hence, it is critical to understand which aspects of RNA base pairing are not captured well by the NNTM.

One known weakness of the current model is the energy function which governs the branching of an RNA secondary structure.
For computational reasons, the entropic cost is modeled as an
affine function with three parameters.
A very natural question to ask is: \textit{How does the optimal secondary structure depend on the branching loop parameters?}

Such a question is an example of a parametric analysis, which illuminates the dependencies of the solution on the underlying optimization parameters. As we illustrate here, methods from geometric combinatorics, specifically polytopes and their normal fans, can be used to answer this question, and others including accuracy and stability. The outline of this paper is as follows. \cref{s:background} contains the basic RNA terminology and  a brief overview of the geometric tools needed for parametric analysis. In~\cref{s:rnapoly} we first give background on the NNTM, specifically the way the multibranch loops are scored. Then we introduce branching polytopes and explain how to compute them using the software \texttt {pmfe} which we have made available online. Finally, we present a few biological questions and discuss mathematical problems related to the branching polytopes they raise. The paper ends with a summary in~\cref{s:conclusions}.

\section{Background}
\label{s:background}

\subsection{Computational molecular biology}
\label{s:compbio}
%% Setting
Nucleotide sequences can be modeled as strings over a finite alphabet. The questions related to DNA and RNA are amenable to analysis using discrete mathematical techniques from algebra, combinatorics, and topology. We refer the reader to the following references for mathematical approaches for studying computational molecular biology \cite{durbin1998,waterman1995,pevzner2000,gusfield1997,pachter2005,bates2005}.
Of particular interest here are techniques from geometric combinatorics, which can inform parametric analysis of models arising from optimization questions, such as DNA sequence alignment and RNA secondary structure prediction.

%% DNA problem
Previous parametric analyses in molecular biology have primarily focused on DNA.  Given sequences of DNA, an important problem is to find regions that have been evolutionary conserved. The first step requires \emph{aligning} the sequences by adding spaces to the sequences until they are the same length. The \emph{DNA sequence alignment} problem is to find the best alignment between sequences given some scoring function \cite{waterman1994}. The optimal solution can be determined via dynamic programming optimization which scores various features and provides global alignment. The optimization fundamentally depends on parameters, and the notion of `best matching' sequences may depend on the value of these parameters. Techniques from geometric combinatorics can be used \cite{waterman1994,dewey2006} to analyze systematically how sequence alignment depends on these parameters. The mathematical background will be presented in \cref{s:polytopes}.

\begin{figure}[htbp]
  \centering
  \includegraphics{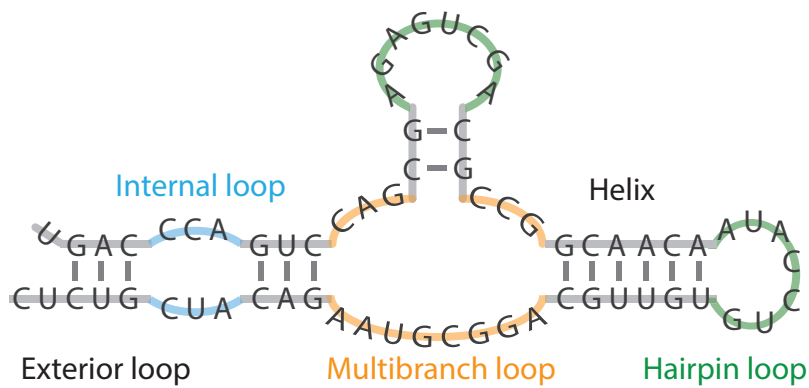}
  \caption{Secondary structure of RNA with substructures.}
  \label{fig:secondary_struct}
\end{figure}

%% Our RNA problem + secondary structures
The problem we focus on here is prediction of the secondary structure of a single given RNA sequence.
Unlike DNA, which typically forms a fully base-paired double helix, RNA typically exists as a single strand whose bases can pair with one another to form diverse and complex secondary structures.
The secondary structure of a given RNA molecule serves as a scaffold for the 3D structure \cite{tinoco1999}.
As shown in \cref{fig:secondary_struct}, an RNA secondary structure consists of stacked base pairs (\emph{helices}) and single-stranded regions (\emph{loops}).
The \emph{degree} of a loop is the number of helices emerging from it (or, equivalently, the number of base pairings in the loop).
\emph{Hairpin loops} have degree 1, \emph{internal loops} have degree 2, and \emph{multibranch loops} have degree greater than 2; each RNA molecule also has a (possibly empty) \emph{exterior loop}, which includes the unpaired bases not contained in other loops.
\Cref{fig:secondary_struct} shows a secondary structure with all of these substructures labeled.

%% Model
The base pairings in the secondary structure are difficult to determine experimentally, necessitating mathematical models to predict the structure \cite{zuker1986}. Indeed secondary structure prediction has been an active area of research for several decades (see survey in \cite{mathews2006}). One approach for predicting secondary structure is calculating the \emph{free thermodynamic energy} of the RNA molecule according to the standard NNTM \cite{mathews1999,turner2009}.
In this model, bases that bond are generally energetically favorable and \emph{stabilize} the RNA with a negative free energy, whereas unpaired bases are energetically disfavorable and \emph{destabilize} the RNA with a positive free energy. Each substructure is assigned an energy value, and the total energy of the secondary structure is determined by summing over its substructures \cite{mathews1999,turner2009}.

%% Since the energy computation decomposes as a sum over independent substructures, dynamic programming can be used to minimize it efficiently.
If we further assume that secondary structure has no crossing base pairs, we could then generate secondary structure of an RNA sequence recursively over its subsequences. This allows us to compute the minimum energy structure efficiently using a dynamic programming algorithm.
This approach is widely studied, and several software packages are available to perform the appropriate computations; widely-used implementations include \texttt{RNAfold} from the Vienna RNA package \cite{hofacker1994}, \texttt{mfold} \cite{zuker1989}, and \texttt{RNAstructure} \cite{reuter2010}.
However, the optimal structure found using the thermodynamic model is often not the correct one. The prediction depends on the thousands of parameter values in the objective function, affecting the substructure prediction. %As a discrete optimization problem, in practice RNA secondary structure is more complicated than DNA sequence alignment because of the increase in scale (additional number of parameters and more recursion calls).

% Parameter problem
Current versions of the scoring model \cite{mathews1999, turner2009} have over eight thousand parameters, but we focus our attention on three related to multibranch loops such as that at the center of \cref{fig:secondary_struct}.
(We will discuss some details of the model and its parameters in \cref{s:nntm}.)
We perform parametric analysis using geometric combinatorics, which has previously been applied to a simplified tree model of RNA base pairing that focuses on branching configurations~\cite{hower2011}, and extend this to study the branching of real RNA sequences under the \textsc{Turner99} NNTM parameters.
In the next section we introduce the required combinatorial and geometric notions to carry out this parametric analysis.

\subsection{Polytopes and geometric combinatorics}
\label{s:polytopes}
%% specifically polytopes and normal fans (meaning and structure related to our project)

The optimization of the NNTM objective function, with an emphasis on branching loop calculations, can be presented as a simple linear programming (LP) problem, and thus well-established mathematical tools can be used to tackle the optimization problem parametrically.
To analyze optimization problems with large sets of feasible solutions, such as all possible secondary structures that could form from a particular sequence of RNA, a (relatively) concise description of the possibilities is required.
This can be achieved by constructing the convex polytope associated with the LP problem at hand.

The theory of convex polytopes and normal fans is well-developed, so we will not reproduce it here. Rather we present just the definitions needed for a rigorous description of the results in \cref{s:rnapoly}.
More information on these applications of polytopes is given in \cite{pachter2005,thomas2006, deloera2012}; readers wishing to gain a thorough understanding of the theory can see \cite{grunbaum2003,ziegler1995,bertsimas1997}.

The theory of linear programming starts from the following question: Given a set of feasible solutions $X$ in $\mathbb{R}_{\geq 0}^n$ and a vector $\vecname{A} \in \mathbb{R}^n$, for which $\vecname{x} \in X$ is the \emph{objective function} $\vecname{A} \vecname{x}$ minimized?

For a fixed $\vecname{A}$, this is a straightforward question; however as $\vecname{A}$ varies, different vectors in the set $X$ will minimize $\vecname{A} \vecname{x}$.
If $X$ is a large or infinite set it may be simpler to deal with a geometric region containing $X$, than with the original set of feasible solutions itself.
A \emph{convex polytope} is a bounded region in $\mathbb{R}^n$ which can be described by a finite number of linear inequalities $\vecname{a}_i \vecname{x} \leq c_i$. A hyperplane is a \emph{supporting hyperplane} for a polytope $P$ if it intersects $P$ non-trivially but no interior point of $P$ is on the hyperplane. A \emph{$k$-face} of $P$ is a $k$-dimensional intersection of $P$ with a  supporting hyperplane. A $0$-face is a \emph{vertex} of $P$, a $1$-face is an \emph{edge}, and so on, with the $n-1$ dimensional faces called \emph{facets}. The (unique) $n$-face is defined to be the entire polytope.

\begin{figure}[htbp]
  \centering
  \begin{tikzpicture}

    \begin{axis}[
      xmin = -2,
      xmax = 11,
      xtick = \empty,
      ymin = -2,
      ymax = 11,
      ytick = \empty,
      axis lines = middle,
      axis line style = {=>, very thick},
      ]

      % Points
      \node [lppoint] (dl) at (axis cs: 4, 3) {};
      \node [lppoint] (ul) at (axis cs: 3, 10) {};
      \node [lppoint] at (axis cs: 4.5, 5) {};
      \node [lppoint] at (axis cs: 5, 7) {};
      \node [lppoint] at (axis cs: 6, 4) {};
      \node [lppoint] at (axis cs: 7, 6) {};
      \node [lppoint] at (axis cs: 3.6, 8.7) {};
      \node [lppoint] at (axis cs: 6.5, 2.4) {};
      \node [lppoint] at (axis cs: 4.7, 6) {};
      \node [lppoint] (ur) at (axis cs: 8, 8) {};
      \node [lppoint] (dr) at (axis cs: 8, 1) {};

      % Named coordinate at origin
      \coordinate (O) at (axis cs: 0, 0) {};

      % Feasible region
      \begin{scope}[on background layer]
        \draw [thick, polyfill] (dl.center) -- (dr.center) -- (ur.center) -- (ul.center) -- cycle;
      \end{scope}
    \end{axis}
  \end{tikzpicture}

  \caption{The convex hull of the set $X$: a polytope with four 0-faces, four 1-faces, and one 2-face.\label{fig:convexhull}}
\end{figure}
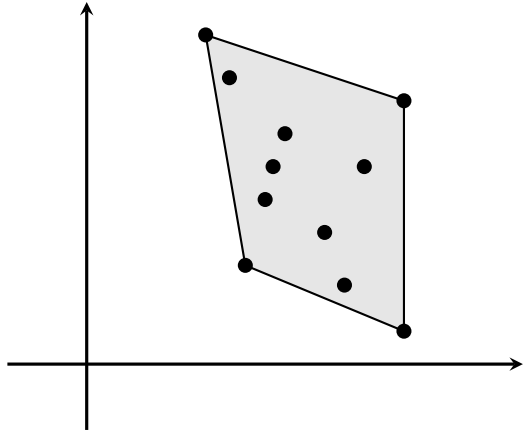

The \emph{convex hull} of a set $X$ is the smallest convex set containing $X$. If $X$ is a finite set, its convex hull $P$ is the set
\begin{equation}
  \label{eq:polydef}
  P = \left\{\sum_{\vecname{x} \in X} b_\vecname{x} \vecname{x}: \text{ each } b_\vecname{x}\geq 0 \text{ and } \sum_{\vecname{x}\in X} b_\vecname{x} = 1 \right\}.
\end{equation}
Equivalently, \cref{eq:polydef} can be used as a definition of a convex polytope---that is, a convex polytope is the convex hull of a finite set.
It is a fundamental theorem in the field that these two definitions of convex polytopes are equivalent.
\Cref{fig:convexhull} shows the convex hull of a finite set in $\mathbb{R}^2$.

The boundary of a polytope is the union of its facets and contains all faces except the $n$-face.
If the polytope $P$ is the convex hull of a set $X$ of feasible solutions, then the boundary of $P$ contains all optimal solutions (i.e. all vectors that minimize $\vecname{A} \vecname{x}$ for non-zero $\vecname{A}$).
Thus, studying the polytope rather than the original set $X$ simplifies the optimization problem.

In practice, however, computing the convex hull of the feasible set $X$ is typically a hard problem.
We usually know too little about the set of feasible solutions to construct the convex hull directly.
In computational biology, it is usually the case that the feasible set is defined implicitly by some intrinsic rules of the linear programming problem.
The real challenge, then, is to extract information about the feasible set from these intrinsic rules.
In this chapter, we consider two different methods for constructing the convex hull: polytope propagation and incremental convex hull.
We will discuss these methods further in \cref{s:polyhedral}.

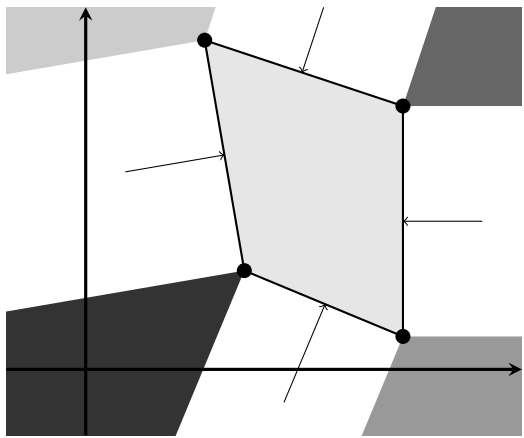
\begin{figure}[htbp]
  \centering
  \begin{tikzpicture}
    \begin{axis}[
      xmin = -2,
      xmax = 11,
      xtick = \empty,
      ymin = -2,
      ymax = 11,
      ytick = \empty,
      axis lines = middle,
      axis line style = {=>, very thick},
      clip=true
      ]

      % Points
      \node [lppoint] (dl) at (axis cs: 4, 3) {};
      \node [lppoint] (ul) at (axis cs: 3, 10) {};
      \node [lppoint] (ur) at (axis cs: 8, 8) {};
      \node [lppoint] (dr) at (axis cs: 8, 1) {};

      % Named coordinate at origin
      \coordinate (O) at (axis cs: 0, 0) {};

      % Feasible region
      \begin{scope}[on background layer]
        \draw [thick, polyfill] (dl.center) -- (dr.center) -- (ur.center) -- (ul.center) -- cycle;
      \end{scope}

      % Normal vectors
      \coordinate (P) at (axis cs:1,6);
      \draw [<-] ($(ul)!(P)!(dl)$) --  (P);
      \coordinate (Q) at (axis cs:5,-1);
      \draw [<-] ($(dl)!(Q)!(dr)$) --  (Q);
      \coordinate (R) at (axis cs:6,11);
      \draw [<-] ($(ul)!(R)!(ur)$) --  (R);
      \coordinate (S) at (axis cs:10,4.5);
      \draw [<-] ($(dr)!(S)!(ur)$) --  (S);

      % Cone vectors
      \fill[cone1] (ur.center) -- ($(ur)!5cm!-90:(ul)$) -- ($(ur)!5cm!90:(dr)$) -- cycle;
      \fill[cone2] (ul.center) -- ($(ul)!5cm!90:(ur)$) -- ($(ul)!5cm!-90:(dl)$) -- cycle;
      \fill[cone3] (dl.center) -- ($(dl)!5cm!90:(ul)$) -- ($(dl)!5cm!-90:(dr)$) -- cycle;
      \fill[cone4] (dr.center) -- ($(dr)!5cm!90:(dl)$) -- ($(dr)!5cm!-90:(ur)$) -- cycle;
    \end{axis}
  \end{tikzpicture}
  \caption{The cones corresponding to the vertices of polytope $P$, drawn ``pointing into'' their vertices. Also shown: the normal vectors to each $1$-face of the polytope.\label{fig:polywithfan}}
\end{figure}

The dual object to the polytope is its \emph{normal fan}.
The normal fan is a collection of regions of $\mathbb{R}^n$ called \emph{cones}.
If $F$ is a face of the polytope, the cone corresponding to $F$, denoted $C_F$, is the set of all vectors $\vecname{B}$ such that any point $\vecname{y} \in F$ minimizes the product $\vecname{B} \vecname{x}$---that is,
\begin{equation*}
  C_F = \{ \vecname{B} \in \mathbb{R}^n: \vecname{By} \leq \vecname{Bx} \text{ for all } \vecname{x} \in P, \vecname{y}\in F\}.
\end{equation*}
The normal fan is represented visually, as in \cref{fig:polywithfan}, by drawing the vectors $\vecname{B}\in C_{F}$ with their heads at the face $F$.

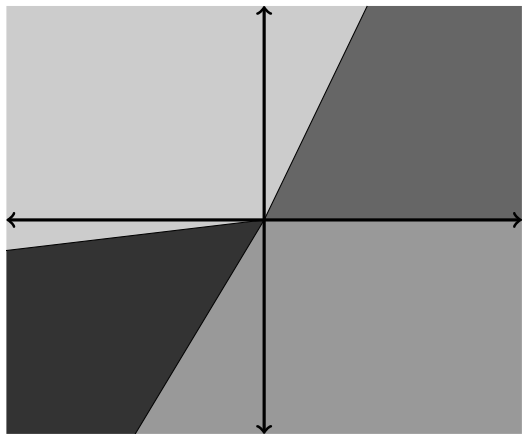
\begin{figure}[htbp]
  \centering
  \begin{tikzpicture}
    \begin{axis}[
      xmin = -2,
      xmax = 2,
      xtick = \empty,
      ymin = -2,
      ymax = 2,
      ytick = \empty,
      axis lines = middle,
      axis line style = {<->, very thick},
      ]

      % Named coordinate at origin
      \coordinate (O) at (axis cs: 0, 0) {};
      \draw[] (O)--(axis cs:-7,-1);
      \draw[] (O)--(axis cs: .8,2);
      \draw[] (O)--(axis cs: 5,0);
      \draw[] (O)--(axis cs: -2,-4);

      \begin{scope}[on background layer]
        \fill[cone1] (O) --(axis cs: 2,0)--(axis cs: 2,2) --(axis cs: .8,2) -- cycle; %ur
        \fill[cone2] (O) -- (axis cs: .8,2) --(axis cs: -2,2) --(axis cs:-2,-.286)-- cycle; %ul
        \fill[cone3] (O) -- (axis cs: -2,-.286) --(axis cs: -2,-2)--(axis cs: -1,-2) -- cycle; %dl
        \fill[cone4] (O)--(axis cs: 2,0)--(axis cs: 2,-2) -- (axis cs:-1,-2)-- cycle; %dr

      \end{scope}

    \end{axis}
  \end{tikzpicture}
  \caption{The normal fan of the polytope fills $\mathbb{R}^2$. \label{fig:normalfan}}
\end{figure}

The normal fan of a polytope of dimension $d$ is a partition of the $\mathbb{R}^d$; every $d$-vector lies in some cone, and the cones are preserved by positive scalar multiplication.  The cones from \cref{fig:polywithfan}, for example, if drawn so that their heads are the origin, clearly fill the entire space $\mathbb{R}^{2}$, as shown in \cref{fig:normalfan}.

Once we have constructed the convex hull of feasible solutions for a particular LP problem, the normal fan provides ways to discuss the stability of a parameter vector $\vecname{A}$.
If $\vecname{A}$ is near the boundary of a cone of the normal fan, then a small change in $\vecname{A}$ will result in a different point of the polytope optimizing the function.
However if $\vecname{A}$ is far from any boundaries of its cone, then small changes will not change which point of the polytope optimizes $\vecname{A} \vecname{x}$.

% Given a convex polytope, constructing its normal fan is an essentially geometric process; it does not require reference to the original LP problem.
% In practice, however, we usually know too little about the set of feasible solutions to construct the convex hull directly.
% In computational biology, it is usually the case that the feasible set is defined implicitly by some intrinsic rules of the linear programming problem.
% The real challenge, then, is to extract information about the feasible set from these intrinsic rules.
% In this chapter, we consider two different methods for constructing the convex hull: polytope propagation and incremental convex hull.
% We will discuss these methods further in \cref{s:polyhedral}.

In \cref{s:rnapoly} we use the NNTM free energy calculation as our objective function and the space of all feasible secondary structures over a given RNA sequence as the point set.
Treating this objective function as a linear functional and applyling the machinery of linear programming produces what we term the \emph{branching polytope}, which we use to analyze the parameters in the model related to multibranch loops.

\section{RNA branching polytope}
\label{s:rnapoly}
\subsection{NNTM}
\label{s:nntm}
In \cref{s:motivation,s:compbio}, we introduced the discrete optimization paradigm for RNA secondary structure prediction.
Before we consider applications of geometric combinatorics for parametric analysis of this method, we review it in greater detail.

Discrete optimization methods for RNA secondary structure prediction derive their objective function from the NNTM.
If we make the standard simplifying assumption to forbid pseudoknots, we can formulate the free-energy minimization problem recursively; dynamic programming can then be used to find the minimum energy and a traceback algorithm can determine a corresponding structure.

The specific objective function used for NNTM analysis has evolved substantially over the years (cf.~\cite{nussinov1978,waterman1978,zuker1981}), with a significant increase in the number of parameters.
The \textsc{Turner99} version considered here, obtained from the Nearest Neighbor Database \cite{turner2009}, has over eight thousand parameters representing the energy contributions of various small substructures---some ($\sim 300$) measured directly through experiments, others ($\sim 7000$ inferred indirectly from experimental data, and finally a handful through machine learning techniques to tune the model.
Among those inferred through machine learning are three connected to multibranch loops, such as the one at the center of \cref{fig:secondary_struct}.

In particular, the \textsc{Turner99} version of the model we study assigns to a given multibranch loop the energy
\begin{equation}
  \label{eq:multibranch-loop}
  \Delta G _{\text{initiation}} = a + b \cdot (\text{\# unpaired nucleotides}) + c \cdot (\text{\# branching helices})
\end{equation}
where $a$, $b$, and $c$ are three of the learned parameters.
The free energy changes associated with multibranch loops have in fact been studied experimentally \cite{diamond2001,mathews2002}.
However, these results do not map neatly onto the linear function from \cref{eq:multibranch-loop}, which is a computational simplification---a logarithmic dependence on loop size would be more biophysically accurate, but the dynamic programming approach requires a recursive decomposition that would not be possible with such a function.
Recent work has led to the development of new approaches to loop energy estimation \cite{zhang2008,aalberts2010}, which rank structures more accurately but do not translate neatly into the discrete optimization setting.
% TODO: say a little more about why \textsc{Turner99} is still used?

Rather than focus on specific values of parameters, we investigate the behavior of the model as a function of the parameters, hoping to gain some insight into (for example) how sensitive the model is to each one and whether certain families of sequences exhibit increased or reduced sensitivity.
If we could compute the collection of all secondary structures $T$ for a given sequence $S$, we could attack this question directly.
Unfortunately, this collection is far too large to compute explicitly---the total number of secondary structures on $n$ bases is known to grow exponentially with $n$ \cite{stein1979}, and specific sequences of biologically-relevant lengths may have $10^{50}$ or more possible structures---so we need to trim the problem down in scope.

To do this, we mathematically reformulate the NNTM as a simple linear functional focusing on the branching loop energy calculations.
For a given RNA sequence, let $T$ be a secondary structure with $x$ multibranch loops, $y$ unpaired nucleotides in those multibranch loops, and $z$ helices branching from those multibranch loops.
Let $a_{99} = 3.4$, $b_{99} = 0$, and $c_{99} = 0.4$ be the values of the parameters $a$, $b$, and $c$ in the \textsc{Turner99} assignment.
Then the energy $\Delta G$ associated to the secondary structure $T$ is given by
\begin{equation}
  \label{eq:def-w}
  \Delta G_{T} = a_{99} x + b_{99} y + c_{99} z + w
\end{equation}
for a ``leftover'' energy value $w$ which can be computed using the model.
We call the vector $\langle x, y, z, w \rangle$ the (branching) \emph{energy signature} of the secondary structure $T$.
Crucially, this energy signature does not actually depend on the \textsc{Turner99} values of $a$, $b$, and $c$; the numbers $x$, $y$, and $z$ are simply the integer counts of certain substructures, while the leftover energy $w$ represents the non-multibranch-loop components of the energy calculation including the energy of the helices that meet the multibranch loops. Treating all other energy components as a single variable $w$ allows us to focus on the multibranch parameters exclusively while also linearizing the optimization problem.

Let $a$, $b$, $c$, and $d$ be four arbitrary rationals.
For a given secondary structure $T$ on a given RNA sequence, we can compute an associated parameterized energy:
\begin{equation}
  \label{eq:energy-linfunc}
  \Delta G_{T} (a, b, c, d)
  = a x + b y + c z + d w
\end{equation}
for  the energy signature $\langle x, y, z, w \rangle$ of $T$.
% (We note for future reference that $\Delta G_{T}$ is a linear function of $\langle a, b, c, d \rangle \in \mathbb{Q}^{4}$.)

Given a particular parameter vector $\langle a, b, c, d \rangle$ and RNA sequence $S$, we can find the secondary structure $T$ which minimizes \cref{eq:energy-linfunc} using existing software; the details of this computation are discussed in \cref{s:pmfe}.
However, to understand how the model behaves for \emph{all possible} values of the parameters $a$, $b$, and $c$, we will need to bring the powerful machinery of linear programming and polyhedral geometry to bear on the problem.

\subsection{Polyhedral methods and convex hulls}
\label{s:polyhedral}

Although the collection of secondary structures for a given sequence is enormous, it is nevertheless finite, and its corresponding energy signatures occupy a bounded region of $\mathbb{R}^{4}$.
We thus shift our attention to this region---specifically, to the \emph{convex hull} of the collection of signature vectors for the sequence which we refer to as the \emph{branching polytope}.
Unlike a classical LP problem, in which the feasible region is explicitly defined by inequalities, the collection of all branching signatures is \emph{implicitly} defined by the NNTM and the RNA sequence.
As a result, we have little information about the branching polytope.
On the other hand, we have a dynamic programming algorithm that can find an energy-minimizing RNA configuration for \emph{any particular} set of parameters.
Hence it is natural to try to construct the convex hull of the signature vectors using this dynamic programming algorithm.

The first approach we consider is polytope propagation \cite{pachter2004}.
Polytope propagation has been used in parametric analysis of various problems, such as DNA sequence alignment and hidden Markov model for gene annotation \cite{pachter2005}. This algorithm computes the convex hull by recursive convex hull and Minkowski sum computations on unions of polytopes. The recursions are related to the dynamic programming algorithm for calculating MFE structures. Namely, the dynamic programming algorithms we use involve decomposing a problem into a sum of smaller problems, each of which is a minimization problem involving only addition.
The minimization operations can be translated into taking convex hulls of unions of polytopes, while the sums can be translated into polytope Minkowski sums, transporting the algorithm directly into the geometric domain.
% This process has an elegant interpretation in terms of tropical geometry which is beyond the scope of this chapter.)
% In another word, we can break down the original minimization problem over the entire sequence into a collection of simpler minimization problems over its subsequences.
% The basic idea of this approach is to start with a small number of known points of the polytope, then use carefully-chosen specializations to iteratively find more points until the entire polytope is constructured.
% Equivalently, we can rewrite the original minimization problem in terms of tropical algebra, that tropical sum $\oplus$ is corresponding to classical minimum and tropical product $\otimes$ is corresponding to classical addition.
% Furthermore, we could `translate' tropical operations into polytope operations.
% The `sum' of polytopes is the convex hull of the union of polytope and the `product' of polytopes is their Minkowski sum.
% This translation yields a recursive representation of the RNA branching polytope.

However, while the translation from dynamic programming to polytope algebra is mathematically very elegant, polytope propagation might not be the most efficient way to compute the RNA branching polytope.
The intermediate polytope computations required in this approach (convex hulls and Minkowski sum) are expensive for complicated polytopes, and these costs quickly add up.
In fact, empirical results show that polytope propagation is often outperformed by incremental
convex hull approaches, especially for high-dimensional models \cite{dewey2006}.
In addition, applying the polytope propagation approach would require re-implementing the optimization algorithm from scratch without taking advantage of the existing fast, peer-reviewed software for NNTM prediction.

As a result, we instead use a variant of the \emph{Beneath-and-Beyond} approach.
The core idea of Beneath-and-Beyond was introduced by Gr\"{u}nbaum in \cite{grunbaum2003} as a method to find the facets of the convex hull of a given point set.
Huggins~\cite{huggins2006} developed  his \texttt{iB4e} algorithm, which applies this idea to specialized LP problems such as ours.
The basic idea of \texttt{iB4e} method is to build the convex hull incrementally---that is, to add one vertex at a time to an already constructed convex hull, by systematically solving LP problems to either expand the hull or confirm that some part of it matches the final hull.
The objective vectors are chosen so that after each iteration, either a new vertex of the convex hull is found or a facet of the convex hull is confirmed.
This ensures that the method requires running the LP solver for no more than $O(V+F)$ objective vectors, if the convex hull has $V$ vertices and $F$ facets.
The algorithm \texttt{iB4e} applies to abstract convex-hull-finding problems in any dimension, so we describe it here in full generality.

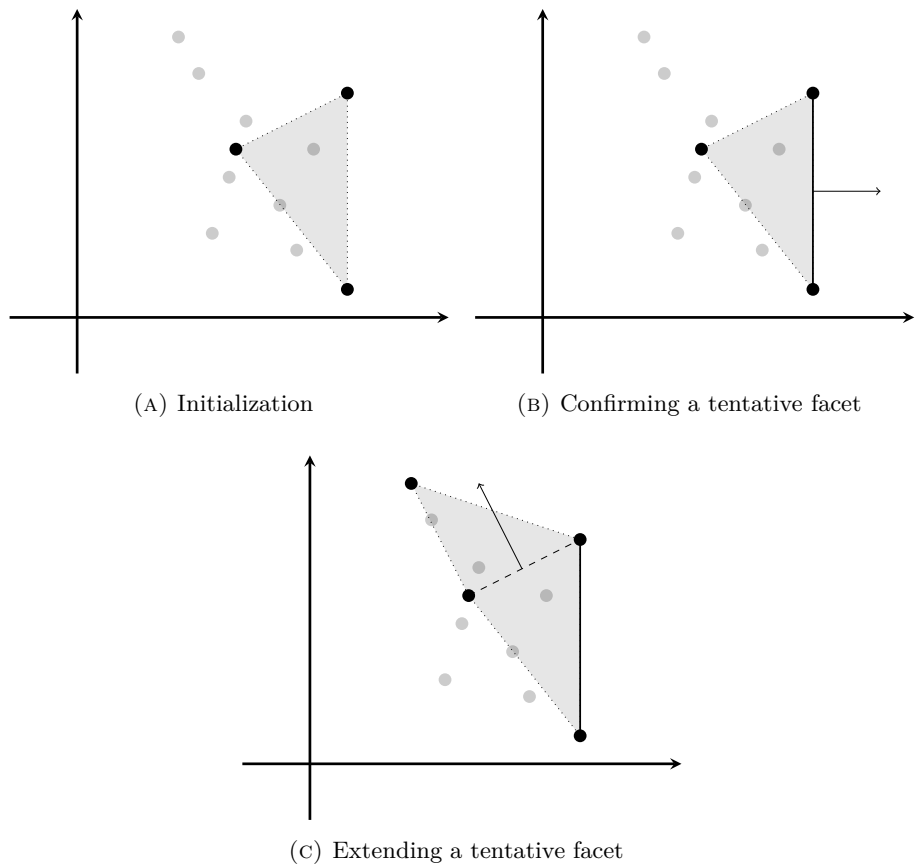
\begin{figure}[hbt]
  \centering
  \newlength{\figwidth}
  \setlength{\figwidth}{0.48\textwidth}
  \begin{subfigure}[t]{\figwidth}
    \resizebox{\figwidth}{!}{
      \begin{tikzpicture}

        \begin{axis}[
          xmin = -2,
          xmax = 11,
          xtick = \empty,
          ymin = -2,
          ymax = 11,
          ytick = \empty,
          axis lines = middle,
          axis line style = {=>, very thick},
          ]

          % Points
          \node [llppoint] (dl) at (axis cs: 4, 3) {};
          \node [llppoint] (ul) at (axis cs: 3, 10) {};
          \node [llppoint] at (axis cs: 4.5, 5) {};
          \node [llppoint] at (axis cs: 5, 7) {};
          \node [llppoint] at (axis cs: 6, 4) {};
          \node [llppoint] at (axis cs: 7, 6) {};
          \node [llppoint] at (axis cs: 3.6, 8.7) {};
          \node [llppoint] at (axis cs: 6.5, 2.4) {};
          \node [lppoint] (pt) at (axis cs: 4.7, 6) {};
          \node [lppoint] (ur) at (axis cs: 8, 8) {};
          \node [lppoint] (dr) at (axis cs: 8, 1) {};

          % Named coordinate at origin
          \coordinate (O) at (axis cs: 0, 0) {};

          % Feasible region
          \begin{scope}[on background layer]
            \draw [dotted, polyfill] (pt.center) -- (dr.center) -- (ur.center) -- cycle;
          \end{scope}
        \end{axis}
      \end{tikzpicture}
    }
    \caption{Initialization\label{fig:bb:init}}
  \end{subfigure}
  \begin{subfigure}[t]{\figwidth}
    \resizebox{\figwidth}{!}{
      \begin{tikzpicture}

        \begin{axis}[
          xmin = -2,
          xmax = 11,
          xtick = \empty,
          ymin = -2,
          ymax = 11,
          ytick = \empty,
          axis lines = middle,
          axis line style = {=>, very thick},
          ]

          % Points
          \node [llppoint] (dl) at (axis cs: 4, 3) {};
          \node [llppoint] (ul) at (axis cs: 3, 10) {};
          \node [llppoint] at (axis cs: 4.5, 5) {};
          \node [llppoint] at (axis cs: 5, 7) {};
          \node [llppoint] at (axis cs: 6, 4) {};
          \node [llppoint] at (axis cs: 7, 6) {};
          \node [llppoint] at (axis cs: 3.6, 8.7) {};
          \node [llppoint] at (axis cs: 6.5, 2.4) {};
          \node [lppoint] (pt) at (axis cs: 4.7, 6) {};
          \node [lppoint] (ur) at (axis cs: 8, 8) {};
          \node [lppoint] (dr) at (axis cs: 8, 1) {};

          % Named coordinate at origin
          \coordinate (O) at (axis cs: 0, 0) {};

          % Feasible region
          \begin{scope}[on background layer]
            \draw [dotted, polyfill] (pt.center) -- (dr.center) -- (ur.center) -- cycle;
            \draw [thick] (dr.center)-- (ur.center);
            \coordinate (S) at (axis cs:10,4.5);
            \draw [->] ($(dr)!(S)!(ur)$) --  (S);
          \end{scope}
        \end{axis}
      \end{tikzpicture}
    }
    \caption{Confirming a tentative facet\label{fig:bb:conf}}
  \end{subfigure}
  \\[\baselineskip]
  \begin{subfigure}[t]{\figwidth}
    \resizebox{\figwidth}{!}{
      \begin{tikzpicture}
        \begin{axis}[
          xmin = -2,
          xmax = 11,
          xtick = \empty,
          ymin = -2,
          ymax = 11,
          ytick = \empty,
          axis lines = middle,
          axis line style = {=>, very thick},
          ]

          % Points
          \node [llppoint] (dl) at (axis cs: 4, 3) {};
          \node [lppoint] (ul) at (axis cs: 3, 10) {};
          \node [llppoint] at (axis cs: 4.5, 5) {};
          \node [llppoint] at (axis cs: 5, 7) {};
          \node [llppoint] at (axis cs: 6, 4) {};
          \node [llppoint] at (axis cs: 7, 6) {};
          \node [llppoint] at (axis cs: 3.6, 8.7) {};
          \node [llppoint] at (axis cs: 6.5, 2.4) {};
          \node [lppoint] (pt) at (axis cs: 4.7, 6) {};
          \node [lppoint] (ur) at (axis cs: 8, 8) {};
          \node [lppoint] (dr) at (axis cs: 8, 1) {};

          % Named coordinate at origin
          \coordinate (O) at (axis cs: 0, 0) {};

          % Feasible region
          \begin{scope}[on background layer]
            \draw [dotted, polyfill] (ul.center)-- (pt.center) -- (dr.center) -- (ur.center) -- cycle;
            \draw [dashed] (pt.center)--(ur.center);
            \draw [thick] (dr.center)-- (ur.center);
            \coordinate (S) at (axis cs:5,10);
            \draw [->] ($(pt)!(S)!(ur)$) --  (S);
          \end{scope}
        \end{axis}
      \end{tikzpicture}
    }
    \caption{Extending a tentative facet\label{fig:bb:ext}}
  \end{subfigure}
  \caption{The Beneath-and-Beyond algorithm for polytope construction\label{fig:bb}}
\end{figure}

Before the main loop of the algorithm can be applied, we must perform an initialization step which finds a collection of points whose affine hull is the same as the affine hull of the branching polytope we aim to construct. (Several approaches to this step are possible; for example in $\mathbb{R}^n$ one might run the LP solver with the $2n$ basis vectors $\left<\pm 1, 0, \dots, 0\right>, \left< 0, \pm 1, \dots,0 \right>, \dots, \left<0, \dots, 0, \pm 1\right>$.)
We then compute the convex hull of these points and label all facets of the convex hull as `tentative,' as illustrated in \cref{fig:bb:init}.

We then begin the main loop of the algorithm. At each step, we pick a tentative facet, use its outer normal vector as objective function, and apply the LP solver with this objective. If the signature of the optimal RNA structure for these parameters is not outside the known hull, as illustrated in \cref{fig:bb:conf}, that facet becomes `confirmed' and the process is restarted with another tentative facet.
However, if the resulting signature is outside the known hull, as illustrated in \cref{fig:bb:ext}, we add it to the vertex set, compute the new convex hull, and label the newly added facets as tentative.

The process is repeated until all facets are confirmed.
In the case of a particular RNA sequence, the set of all possible secondary structures is finite and thus set of signatures is also finite.  The process must terminate and its end result is the convex hull of the set of \emph{all} solutions to the LP solver.
This is the branching polytope, and once we have computed it we can easily compute its normal fan using standard methods and proceed to study the model.

\subsection{Computing the branching polytope with \texttt{pmfe}}
\label{s:pmfe}
It is conceptually natural to apply the Beneath-and-Beyond algorithm to compute the branching polytope of a particular RNA sequence by treating the parametrized NNTM minimization problem as a linear program.
However, coupling the two algorithms to perform the computation requires some care.
We introduce the software \texttt{pmfe}, available publicly at \url{https://github.com/AMS-MRC-disc-math-bio/pmfe}.
Instructions to build and install the software are provided in the file \url{README.md} of that repository.
Alternatively, a ready-to-use version of the software is available in a Docker container, posted publicly on the Docker hub as \url{agdphd/pmfe}.

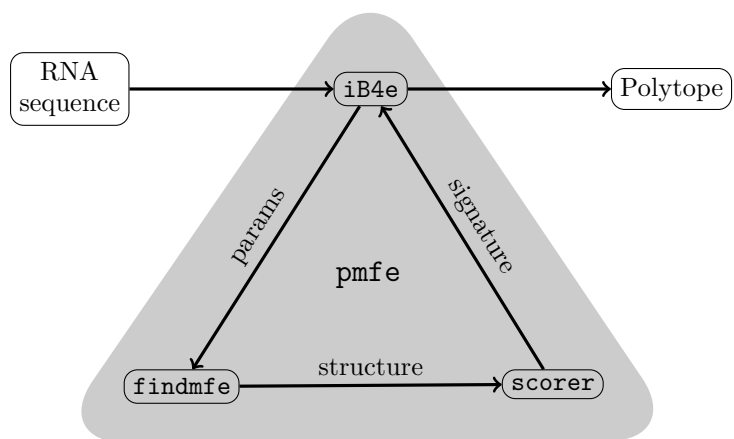
\begin{figure}[htbp]
  \centering
  \begin{tikzpicture}[auto, node distance = 4cm]
    \node [fcentry] (RNA) {RNA\\sequence};
    \node [fcentry, right of=RNA] (iB4e) {\texttt{iB4e}};
    \node [fcentry, below left = 3.5cm and 1.25cm of iB4e] (findmfe) {\texttt{findmfe}};
    \node [fcentry, below right = 3.5cm and 1.25cm of iB4e] (scorer) {\texttt{scorer}};
    \node [fcentry, right of=iB4e] (poly) {Polytope};

    \node (pmfe) at (barycentric cs:iB4e=1.2,findmfe=1,scorer=1) {\Large\texttt{pmfe}};
    \fill [fcwrapper] ([yshift=1.5cm] iB4e.center) -- ([xshift=-1.75cm, yshift=-.75cm] findmfe.center) -- ([xshift=1.75cm, yshift=-.75cm] scorer.center) -- cycle;

    \path [fcarrow] (iB4e) edge node [above, sloped] {params} (findmfe);
    \path [fcarrow] (findmfe) edge node {structure} (scorer);
    \path [fcarrow] (scorer) edge node [above, sloped] {signature} (iB4e);
    \path [fcarrow] (RNA) edge (iB4e);
    \path [fcarrow] (iB4e) edge (poly);
  \end{tikzpicture}
  \caption{Control flow of the \texttt{pmfe} software\label{fig:pmfe-flow}}
\end{figure}

The \texttt{pmfe} package has three components (see \cref{fig:pmfe-flow}): \texttt{iB4e}, \texttt{findmfe}, and \texttt{scorer}.
\begin{itemize}
\item
  \texttt{iB4e} is an implementation of Huggins' algorithm as a header-only template library in C++, taking advantage of the CGAL computational geometry library \cite{cgal} to handle the geometric computations.

\item
  \texttt{findmfe} takes a parameter vector $\vecname{P} = \langle a, b, c, d \rangle$ and an RNA sequence $S$ as input and returns as output a secondary structure on $S$ which has minimal energy with respect to $\vecname{P}$.
  From the available software packages which implement NNTM optimization, we chose to base \texttt{findmfe} on \texttt{GTfold} \cite{mathuriya2009} because it is parallelized and thus able to take advantage of modern multi-core desktop computers.
  Some modifications were required to adapt it to our use.
  For computational efficiency, \texttt{GTfold} performs all arithmetic with fixed precision of two decimal places; we converted it to use arbitrary-precision rational arithmetic based on the GMP library \cite{gmp}, which increases running time by an order of magnitude but ensures that the results are exact even when parameters are not multiples of $1/100$.
  We also added an interface to modify the multibranch parameters $\langle a, b, c \rangle$ and the dummy scaling parameter $d$ programmatically, allowing the Beneath-and-Beyond algorithm to treat \texttt{GTfold} as a function of these four variables and call it repeatedly.

\item
  \texttt{scorer} takes an RNA sequence $S$ and a secondary structure $T$ and returns the signature vector $\langle x, y, z, w \rangle$ of $T$ according to \cref{eq:def-w}.
  (In particular, it computes the value of $w$ by computing the energy decomposition of $T$ under the \textsc{Turner99} parameters.)
\end{itemize}

We combine these three components with the control flow illustrated in \cref{fig:pmfe-flow} to create the \texttt{pmfe} software package.
Given an RNA sequence $S$, the \texttt{iB4e} module repeatedly generates parameter vectors according to Huggins' algorithm.
Each of these vectors is sent to the \texttt{findmfe} module as a query, which generates a secondary structure with minimal energy.
Each such secondary structure is sent to the \texttt{scorer} module to find its signature vector, which is then sent back to \texttt{iB4e} as the response to the query.
After some number of iterations of this loop, the convex hull of these signature points $s$ is found to be equal to the branching polytope, which is returned to the user.
(For convenience, we in fact provide the user with a file containing both the signatures $s$ and their associated structures $T$ in condensed dot-bracket notation.)

Since the software \texttt{pmfe} which we introduce is complex and extensively modifies the peer-reviewed codebase of \texttt{GTfold}, we provide a testing framework to ensure correctness.
Tests are available to ensure that \texttt{pmfe} returns the same structures and scores as \texttt{GTfold}, using both the classical \textsc{Turner99} parameters \cite{mathews1999} and various modified parameter sets for a variety of natural and synthetic RNA sequences.
Details for running these tests are available in the \texttt{pmfe} repository.

\begin{figure}[htbp]
  \centering
  \vspace{3eX} % The figure has a bad bounding box for some reason. This is a quick, easy fix. --AGD
  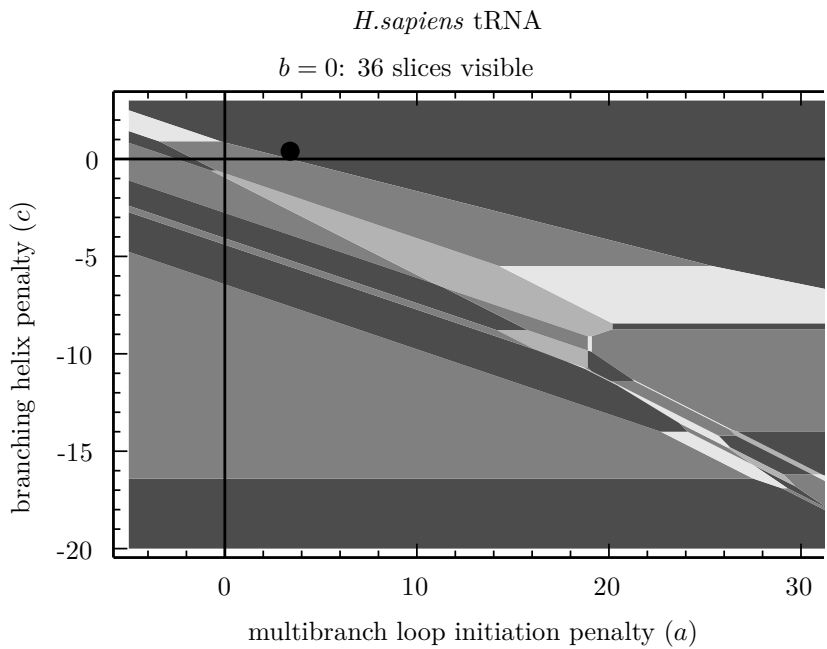
  \vspace{3eX} % The figure has a bad bounding box for some reason. This is a quick, easy fix. --AGD
  \caption{The intersection of the parameter space decomposition for a \textit{H.sapiens} tRNA sequence with $b = 0$.}
  \label{fig:hsapiens_tRNA}
\end{figure}

We also provide a module \texttt{rna\textunderscore{}poly} in the SageMath computer algebra system \cite{sage} which can be used to study the branching polytopes produced by \texttt{pmfe}.
This can be used to generate a variety of visualizations of the normal fan of the branching polytope.
In particular, it provides a simple interface for taking the $d = 1$ slice of the fan, giving a polyhedral partition of $\mathbb{R}^{3}$ whose regions are collections of parameters $\langle a, b, c \rangle$ which yield the same secondary structure for the sequence under study.
The results are illustrated in \cref{fig:hsapiens_tRNA}, which shows the $b = 0, d = 1$ slice of the normal fan of the branching polytope for a \textit{H.sapiens} tRNA sequence.
The \textsc{Turner99} parameter values $a_{99} = 3.4, c_{99} = 0.4$ are marked with a circle just northeast of the origin.

\subsection{Biological questions leading to mathematical problems}
\label{s:bio-probs}
The RNA branching polytope and the associated polyhedral decomposition of the three-dimensional multibranch parameter space  obtained by intersecting its normal fan with $d=1$ encapsulate the dependency of the optimization on the parameters $a, b, c$.
Various questions related to accuracy and stability can be addressed through the analysis of the polytope and the polyhedral decomposition.
However, answering such questions in a manner that is  biologically relevant requires consideration of certain subtleties and this leads to interesting mathematical problems.
Here we discuss a few biological questions that could be addressed through the parametric analysis and related mathematical problems that arise.

\textbf{Question 1:} \emph{How robust is the optimization for a given RNA sequence?}

In the strictest sense, the prediction is sensitive if a variation of the parameters within an allowed margin of error produces different optimal secondary structures. This occurs if the distance of the \textsc{Turner99} parameters $(a_{99}, b_{99}, c_{99})$ from the boundary of the polyhedron that contains this point is smaller than the allowed margin of error. For example, \cref{fig:hsapiens_tRNA} shows the position of $(a_{99}, b_{99}, c_{99})$ within the polyhedral decomposition of a \textit{H.sapiens} tRNA (only the $b=0$ slice is depicted, which is the baseline value in \textsc{Turner99}).
Its distance to the boundary is $0.13699$ and, for instance, optimization using $a= 3.39130, b=-0.14786, c=0.37391$ yields structures with a different signature.

However, this by itself does not mean that the optimization is very sensitive; analysis of the structure space is required so that one can quantify the sensitivity.
Namely, two structures with different signatures may still have a lot of structural similarities (for example, very similar long helices and branching pattern), in which case one would not necessarily say that the prediction is very sensitive.
Moreover, since the structure-to-signature map is not one-to-one, when quantifying the structural differences in the optimal structures one actually needs to compare two sets of structures, each one corresponding to a given signature.
In short, in the assessment of robustness we face the following problem.

\textbf{Problem 1:} \emph{Define a good representative of the set of structures that correspond to the same signature.}

The problem of representing a whole set of structures by a single one has appeared before, for example in the context of compact representation of a sample of structures.
However, the methods developed for this (e.g.~sfold clustering \cite{ding2005} and profiling \cite{rogers2014}) usually assume structural similarities of the elements in the set.
Here, though, we are presented with a different set-up: the structures mapping to the same signature do not a priori have any common motifs.
Therefore, choosing one consensus structure as the other methods do may not  yield a reasonable representative of the whole set.

\textbf{Question 2:} \emph{How much can the prediction of secondary structure be improved for a given sequence?}

For sequences for which the native structure has been obtained experimentally or via comparative sequence analysis, one could assess the accuracy of the NNTM  by comparing it to the MFE structure.
In fact, since the structures corresponding to all signatures on the boundary of the branching polytope can in practice be obtained fairly efficiently \cite{wuchty1999}, one can precisely determine branching parameters that would yield a signature corresponding to a structure that is closest to the native one.
However, here we again face the problem that besides the most accurate one, that signature corresponds to a whole set of MFE structures that might be very different from the native structure.
In such a case, the almost accurate structure might be unrecognizable. Therefore, in assessing accuracy, the problem of finding a good representative structure for a given signature is still very relevant.
Furthermore, even if we can solve this problem, the most accurate prediction obtained this way may require a triple of parameters $(a,b,c)$ that is very different from $(a_{99}, b_{99}, c_{99})$, which may not be very desirable.
Namely, even though $(a_{99}, b_{99}, c_{99})$ were not obtained experimentally, they still might have some biological relevance since the initiation point for the genetic algorithm used was suggested by stabilities determined by optical melting for an RNA multibranch loop with three branching helices \cite{mathews1999}.
This suggests the following natural problem.

\textbf{Problem 2:} \emph{Maximize the improvement in accuracy while minimizing parameter change.}

One mathematical subproblem here is to formulate an objective function which appropriately measures the difference between two RNA secondary structures.
While there are simple structure metrics that can be very efficiently computed (e.g. the base pair metric based on symmetric set difference), other  metrics may be more appropriate to quantify the structural differences \cite{moulton2000}.

Another important subproblem is formulating an appropriate metric to measure changes in parameter values.
Standard metrics such as the $p$-metrics can be applied, since the parameter vectors are taken from $\mathbb{R}^{4}$, but it is not clear which is most appropriate.
Furthermore, careful consideration should be given to how each component of the parameter vector is weighted: a unit change in each coordinate may have significantly different impacts on the secondary structure. For example, since a given multibranch loop in a structure has several unpaired nucleotides, a small change (e.g. by $0.1$) in $b$ in general has bigger effects on the prediction than the same change in $a$.

\textbf{Question 3:} \emph{Is there a choice of branching parameters that would improve the prediction accuracy for a family of sequences?}

\begin{figure}[htbp]
  \centering
  \vspace{3eX} % The figure has a bad bounding box for some reason. This is a quick, easy fix. --AGD
  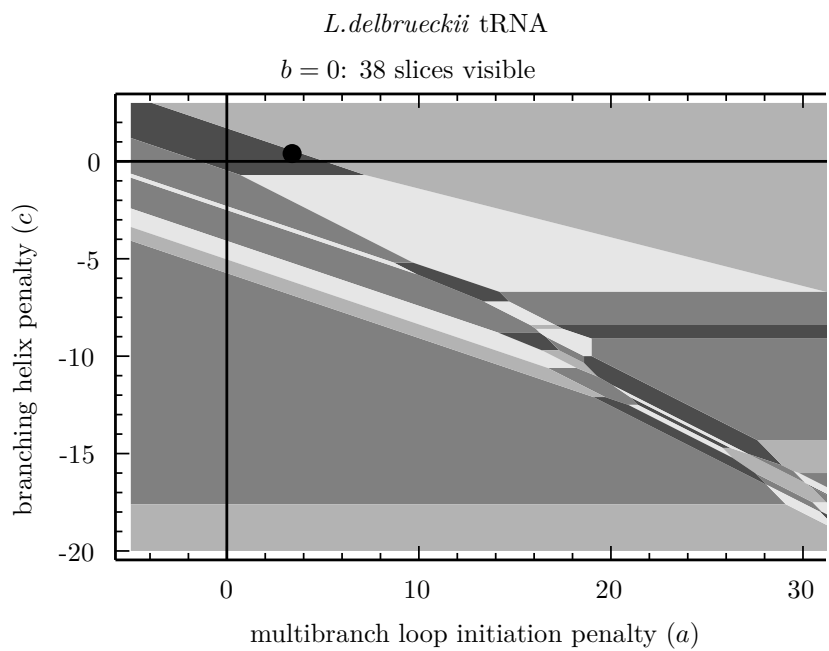
  \vspace{3eX} % The figure has a bad bounding box for some reason. This is a quick, easy fix. --AGD
  \caption{The $b = 0$ slice of the parameter space decomposition for a \textit{L.delbrueckii} tRNA sequence.}
  \label{fig:ldelbrueckii_tRNA}
\end{figure}

Homologous sequences perform the same function in different species and they fold into essentially the same structure.
However, their MFE prediction accuracies can vary significantly. For instance, within the tRNA family, the accuracy calculated as the F-measure (the harmonic mean of the MFE sensitivity and positive predictive value against true positive base pairs) ranges from the minimal $0$ to the maximal $1$ \cite{rogers2014}.
Hence, it is natural to try to find a set of parameters which yields the highest accuracy for a family of sequences.
For this, one needs to address the question: \emph{Are there any patterns in the dependency of the prediction on the parameters across a family of homologous sequences?}
Mathematically, one could phrase this as a problem of finding similarities between polytopes in $\mathbb{R}^4$ or between polyhedral decompositions of $\mathbb{R}^3$.
For example, \cref{fig:ldelbrueckii_tRNA} shows the $b=0$ slice of the polyhedral decompositions for a tRNA sequence of \textit{L.delbrueckii}.
By visually comparing this figure to \cref{fig:hsapiens_tRNA} one can notice similarities in the directions of the splitting hyperplanes.
The challenge of course is to quantify such similarities. So, we face the following problem.

\textbf{Problem 3:} \emph{Find a way to compare branching polytopes.}

Comparison of polytopes is a problem that has appeared before, for instance in relation to  the traveling salesman problem in optimization \cite{goemans1995}.
Even in optimization problems when one works with relaxation polytopes that are, loosely speaking, ``close,'' comparing relaxations is not well understood and different comparison methods have been suggested (e.g.~\cite{lee1994}).
In our case, an additional challenge is the fact that the polytopes to be compared do not a priori satisfy any mathematically tractable assumptions for closeness.

\section{Conclusions}
\label{s:conclusions}

The affine energy function which governs the branching of an RNA secondary structure under the NNTM optimization is a known weakness of the thermodynamic model, and the methods from geometric combinatorics outlined here offer significant potential to assess its accuracy and stability through a parametric analysis.  While simplified models are certainly more tractable, the new \texttt{pmfe} computational framework makes possible the analysis of branching polytopes for real RNA sequences for the first time.  As illustrated by the qualitative similarities between \cref{fig:hsapiens_tRNA} and  \cref{fig:ldelbrueckii_tRNA}, this approach is capturing some interesting characteristics which are preserved between two different sequences.  Moreover, as discussed, moving beyond the qualitative comparison of polytope slices to quantify their similarities and differences presents some interesting mathematical challenges.  Consequently, it remains to be seen whether the observed patterns in these proof-of-principle results are resulting from the structure of the thermodynamic optimization and/or the coding of structural motifs in the base pairing of these RNA sequences.  Resolving this dichotomy will reveal much about the interplay between mathematics and biology in the prediction of RNA secondary structures.

\bibliographystyle{amsplain}
\bibliography{chapter-rna-branching}

\providecommand{\bysame}{\leavevmode\hbox to3em{\hrulefill}\thinspace}
\providecommand{\MR}{\relax\ifhmode\unskip\space\fi MR }
\providecommand{\MRhref}[2]{%
  \href{http://www.ams.org/mathscinet-getitem?mr=#1}{#2}
}
\providecommand{\href}[2]{#2}
\begin{thebibliography}{10}

\bibitem{aalberts2010}
Daniel~P Aalberts and Nagarajan Nandagopal, \emph{A two-length-scale polymer
  theory for {RNA} loop free energies and helix stacking}, RNA \textbf{16}
  (2010), no.~7, 1350--1355.

\bibitem{bates2005}
Andrew~D Bates and Anthony Maxwell, \emph{{DNA} topology}, Oxford University
  Press, Oxford, UK, 2005.

\bibitem{bertsimas1997}
Dimitris Bertsimas and John~N Tsitsiklis, \emph{Introduction to linear
  optimization}, Athena Scientific Series in Optimization and Neural
  Computation, no.~6, Athena Scientific, 1997.

\bibitem{deloera2012}
Jes{\'{u}}s De~Loera, Raymond Hemmecke, and Matthias K{\"{o}}eppe,
  \emph{Algebraic and geometric ideas in the theory of discrete optimization},
  MOS-SIAM Series on Optimization, Society for Industrial and Applied
  Mathematics, Philadelphia, PA, USA, 2012.

\bibitem{dewey2006}
Colin~N Dewey, Peter~M Huggins, Kevin Woods, Bernd Sturmfels, and Lior Pachter,
  \emph{Parametric alignment of {D}rosophila genomes}, PLoS Comput Biol
  \textbf{2} (2006), no.~6, e73.

\bibitem{diamond2001}
Joshua~M Diamond, Douglas~H Turner, and David~H Mathews, \emph{Thermodynamics
  of three-way multibranch loops in {RNA}}, Biochemistry \textbf{40} (2001),
  no.~23, 6971--6981.

\bibitem{ding2005}
Ye~Ding, Chi~Yu Chan, and Charles~E Lawrence, \emph{{RNA} secondary structure
  prediction by centroids in a {B}oltzmann weighted ensemble}, RNA \textbf{11}
  (2005), no.~8, 1157--1166.

\bibitem{durbin1998}
Richard Durbin, Sean~R Eddy, Anders Krogh, and Graeme Mitchison,
  \emph{Biological sequence analysis: Probabilistic models of proteins and
  nucleic acids}, Cambridge University Press, 1998.

\bibitem{goemans1995}
Michel~X Goemans, \emph{Worst-case comparison of valid inequalities for the
  {TSP}}, Mathematical Programming \textbf{69} (1995), no.~1-3, 335--349.

\bibitem{gmp}
Torbj{\"o}rn Granlund and {the GMP development team}, \emph{{GNU MP}: {T}he
  {GNU} {M}ultiple {P}recision {A}rithmetic {L}ibrary}, 6.0.0a ed., 2014,
  \url{http://gmplib.org/}.

\bibitem{grunbaum2003}
Branko Gr{\"u}nbaum, \emph{Convex polytopes}, Graduate Texts in Mathematics,
  vol. 221, Springer-Verlag, New York, 2003.

\bibitem{gusfield1997}
Dan Gusfield, \emph{Algorithms on strings, trees and sequences: computer
  science and computational biology}, Cambridge University Press, 1997.

\bibitem{hofacker1994}
Ivo~L Hofacker, Walter Fontana, Peter~F Stadler, L~Sebastian Bonhoeffer,
  Manfred Tacker, and Peter Schuster, \emph{Fast folding and comparison of
  {RNA} secondary structures}, Monatshefte f{\"u}r Chemie/Chemical Monthly
  \textbf{125} (1994), no.~2, 167--188.

\bibitem{hower2011}
Valerie Hower and Christine~E Heitsch, \emph{Parametric analysis of {RNA}
  branching configurations}, Bulletin of Mathematical Biology \textbf{73}
  (2011), no.~4, 754--776.

\bibitem{huggins2006}
Peter Huggins, \emph{{\textit{iB4e}}: A software framework for parametrizing
  specialized {LP} problems}, Mathematical Software--ICMS 2006, Springer Berlin
  Heidelberg, 2006, pp.~245--247.

\bibitem{lee1994}
Jon Lee and Walter~D Morris, \emph{Geometric comparison of combinatorial
  polytopes}, Discrete Applied Mathematics \textbf{55} (1994), no.~2, 163--182.

\bibitem{mathews1999}
David~H Mathews, Jeffrey Sabina, Michael Zuker, and Douglas~H Turner,
  \emph{Expanded sequence dependence of thermodynamic parameters improves
  prediction of {RNA} secondary structure}, Journal of Molecular Biology
  \textbf{288} (1999), no.~5, 911--940.

\bibitem{mathews2002}
David~H Mathews and Douglas~H Turner, \emph{Experimentally derived
  nearest-neighbor parameters for the stability of {RNA} three- and four-way
  multibranch loops}, Biochemistry \textbf{41} (2002), no.~3, 869--80.

\bibitem{mathews2006}
\bysame, \emph{Prediction of {RNA} secondary structure by free energy
  minimization}, Current Opinion in Structural Biology \textbf{16} (2006),
  no.~3, 270--278.

\bibitem{mathuriya2009}
Amrita Mathuriya, David~A Bader, Christine~E Heitsch, and Stephen~C Harvey,
  \emph{{GTfold}: a scalable multicore code for {RNA} secondary structure
  prediction}, Proceedings of the 2009 ACM symposium on Applied Computing, SAC
  '09, ACM, 2009, pp.~981--988.

\bibitem{moulton2000}
Vincent Moulton, Michael Zuker, Michael Steel, Robin Pointon, and David Penny,
  \emph{Metrics on {RNA} secondary structures}, Journal of Computational
  Biology \textbf{7} (2000), no.~1-2, 277--292.

\bibitem{nussinov1978}
Ruth Nussinov, George Pieczenik, Jerrold~R Griggs, and Daniel~J Kleitman,
  \emph{Algorithms for loop matchings}, SIAM Journal on Applied mathematics
  \textbf{35} (1978), no.~1, 68--82.

\bibitem{pachter2004}
Lior Pachter and Bernd Sturmfels, \emph{Parametric inference for biological
  sequence analysis}, Proceedings of the National Academy of Sciences of the
  United States of America \textbf{101} (2004), no.~46, 16138--16143.

\bibitem{pachter2005}
\bysame, \emph{Algebraic statistics for computational biology}, vol.~13,
  Cambridge University Press, 2005.

\bibitem{pevzner2000}
Pavel Pevzner, \emph{Computational molecular biology: an algorithmic approach},
  MIT Press, 2000.

\bibitem{reuter2010}
Jessica~S Reuter and David~H Mathews, \emph{{RNAstructure}: software for {RNA}
  secondary structure prediction and analysis}, BMC bioinformatics \textbf{11}
  (2010), no.~1, 129.

\bibitem{rogers2014}
Emily Rogers and Christine~E Heitsch, \emph{Profiling small {RNA} reveals
  multimodal substructural signals in a {B}oltzmann ensemble}, Nucleic Acids
  Research (2014).

\bibitem{stein1979}
PR~Stein and MS~Waterman, \emph{On some new sequences generalizing the
  {C}atalan and {M}otzkin numbers}, Discrete Mathematics \textbf{26} (1979),
  no.~3, 261--272.

\bibitem{sage}
W.\thinspace{}A. Stein et~al., \emph{{S}age {M}athematics {S}oftware ({V}ersion
  6.7)}, The Sage Development Team, 2015, {\tt http://www.sagemath.org}.

\bibitem{cgal}
{The CGAL Project}, \emph{{CGAL} user and reference manual}, {4.6} ed., 2015.

\bibitem{thomas2006}
Rekha~R Thomas, \emph{Lectures in geometric combinatorics}, vol.~33, American
  Mathematical Society, 2006.

\bibitem{tinoco1999}
Ignacio Tinoco and Carlos Bustamante, \emph{How {RNA} folds}, Journal of
  Molecular Biology \textbf{293} (1999), no.~2, 271--281.

\bibitem{turner2009}
Douglas~H Turner and David~H Mathews, \emph{{NNDB}: the nearest neighbor
  parameter database for predicting stability of nucleic acid secondary
  structure}, Nucleic Acids Research (2009), gkp892.

\bibitem{waterman1978}
Michael Waterman, \emph{Secondary structure of single-stranded nucleic acids},
  Studies on foundations and combinatorics, Advances in mathematics:
  supplementary studies, Academic Press, 1978, pp.~167--212.

\bibitem{waterman1994}
Michael~S Waterman, \emph{Parametric and ensemble sequence alignment
  algorithms}, Bulletin of Mathematical biology \textbf{56} (1994), no.~4,
  743--767.

\bibitem{waterman1995}
\bysame, \emph{Introduction to computational biology: Maps, sequences and
  genomes}, CRC Press, 1995.

\bibitem{wuchty1999}
Stefan Wuchty, Walter Fontana, Ivo~L Hofacker, Peter Schuster, et~al.,
  \emph{Complete suboptimal folding of {RNA} and the stability of secondary
  structures}, Biopolymers \textbf{49} (1999), no.~2, 145--165.

\bibitem{zhang2008}
Jian Zhang, Ming Lin, Rong Chen, Wei Wang, and Jie Liang, \emph{Discrete state
  model and accurate estimation of loop entropy of {RNA} secondary structures},
  The Journal of chemical physics \textbf{128} (2008), no.~12, 125107.

\bibitem{ziegler1995}
G{\"{u}}nter~M Ziegler, \emph{Lectures on polytopes}, Graduate Texts in
  Mathematics, Springer-Verlag, 1995.

\bibitem{zuker1986}
Michael Zuker, \emph{{RNA} folding prediction: The continued need for
  interaction between biologists and mathematicians}, Some mathematical
  questions in biology: {DNA} sequence analysis, Lectures on Mathematics in the
  Life Sciences, vol.~17, American Mathematical Society, 1986, pp.~87--124.

\bibitem{zuker1989}
\bysame, \emph{On finding all suboptimal foldings of an {RNA} molecule},
  Science \textbf{244} (1989), no.~4900, 48--52.

\bibitem{zuker1981}
Michael Zuker and Patrick Stiegler, \emph{Optimal computer folding of large
  {RNA} sequences using thermodynamics and auxiliary information}, Nucleic
  Acids Research \textbf{9} (1981), no.~1, 133--148.

\end{thebibliography}
\end{document}